# Asymmetry in the effect of magnetic field on photon detection and dark counts in bended nanostrips

A. Semenov[1], I. Charaev[2], R. Lusche[1], K. Ilin[2], M. Siegel[2], H.-W. Hübers[1,3], N. Bralović[4], K. Dopf[4], and D.Yu. Vodolazov[5]

[1] Institute of Optical Sensor Systems, German Aerospace Center (DLR), Rutherfordstrasse 29, 12489 Berlin Germany

[2] Institute of Micro- and Nanoelectronic Systems, Karlsruhe Institute of Technology (KIT), Hertzstrasse 16, 76187 Karlsruhe, Germany

[3] Humboldt-Universität zu Berlin, Department of Physics, Newtonstraße 15, 12489 Berlin, Germany

[4] Light Technology Institute (LTI), Karlsruhe Institute of Technology (KIT), Kaiserstrasse 12, 76131 Karlsruhe, Germany

[5] Institute of Physics of Microstructures, Russian Academy of Sciences, 603950, Nizhny Novgorod, GSP-105, Russia & Lobachevsky State University of Nizhny Novgorod, 23 Gagarin Avenue, 603950 Nizhny Novgorod, Russia

Keywords: bended superconducting nanostrips, current crowding, photon detection, dark counts, magnetic field asymmetry

PACS number(s): 74.25.Ha, 74.78.-w, 85.25.Oj, 78.67.Uh

Current crowding in the bends of superconducting nano-structures not only restricts measurable critical current in such structures but also redistributes local probabilities for dark and light counts to appear. Using structures from strips in the form of a square spiral which contain bends with the very same curvature with respect to the directions of bias current and external magnetic field, we have shown that dark counts as well as light counts at small photon energies originate from areas around the bends. The minimum in the rate of dark counts reproduces the asymmetry of the maximum critical current density as function of the magnetic field. Contrary, the minimum in the rate of light counts demonstrate opposite asymmetry. The rate of light counts become symmetric at large currents and fields. Comparing locally computed absorption probabilities for photons and the simulated threshold detection current we found the approximate locations of areas near bends which deliver asymmetric light counts. Any asymmetry is absent in Archimedean spiral structures without bends.

# I. INTRODUCTION

Meandered nanometer-wide superconducting strips are commonly used for detection of single photons in the near infrared spectral range [1]. Besides the efficiency of photon detection statistical fluctuations in the form of dark counts restrict the minimum detectable photon flux. Recently it has become clear that in stripes with bends current crowding limits the achievable supercurrent to a value noticeably lower than the depairing current in straight fragments of strips [2]. Everywhere at edges, where current rounds a sharp corner, local current density increases which causes a local reduction of the free energy barrier for nucleation of magnetic vortices. Among different topological fluctuations, hopping of vortices across the strip is commonly considered as mechanism of dark counts [3-5]. Hence, sharp turns become preferable places where dark counts may originate from. Only a few indirect experimental verifications of this suggestion have been reported. Engel et al [5] found a slight asymmetry in the rate of dark counts in magnetic field and assigned it to geometric differences in right and left turns in their meander structure. Akhlaghi et al. [6] have shown that, in a nanowire with a single bend, rounding the sharp inner corner of the bend results in an increase of the critical current and in a reduction of the dark count rate of the whole structure. Lusche et al [7] found differences in current dependencies of the vortex energy-barrier in case of light and dark counts and associated them with different locations of these events.

Light counts in nanowires are also related to either current-assisted or fluctuation-assisted vortex crossing. In the first deterministic scenario, a photon creates a hot-spot in the strip which forces the current density to redistribute around the absorption site. A vortex nucleates at any point where after current redistribution the velocity of the superconducting condensate locally achieves its critical value [8]. This might be either a single vortex near the strip edge or a vortex-antivortex pair (VAP) close to the midline of the strip. The vortex is then swept by the Lorentz force across the strip. The energy dissipated along the trajectory of the vortex in the strip cause the formation of a normal belt. In the latter statistical scenario, a vortex crosses with certain thermodynamic probability the entire strip through the segment where the energy barrier is reduced due to photon absorption [9, 10]. Discovering the local nature of count events has made it possible to bridge between these two scenarios in the framework of the deterministic model. Studies of the effect of the external magnetic field on the light count rate [5, 7] have shown that the energy barrier differently depends on the current for low and high energy photons and that the variation of the barrier with the photon energy deviates noticeably from the model predictions for straight strips [9]. These inconsistencies are partly due to simplifications of the boundary conditions in the model of Ref. 9. They could be partly relaxed by suggesting different locations of light counts for photons with different energies [7].

However, the meander structure itself prevents one from figuring out where locally count events originate from. Differentiating straights portions and corners by applying a magnetic field runs into the problem that a meander has bends with opposite symmetry with respect to the direction of current flow and magnetic field. The dual symmetry of the structure masks the expected asymmetry in magnetic fields for count events which occur in bends. Using single bends and bridges helps solving the problem but has its own complication such as resonance effects in the absorption probability for particular wavelengths and current crowding imposed by closely spaced contacts. Furthermore, optical coupling to small structures is deteriorated. In this work, we studied specially designed square-shaped spiral structures which contain bends with only one symmetry with respect to current and magnetic field and have an optical coupling efficiency comparable to the meander structures. As a reference, we used spiral structures without bends. We show that, in accordance to a common understanding of the current crowding effect in magnetic fields [11-13], the magnetic field dependences of the critical current in square spirals is non-symmetric and that this asymmetry is reversible with either current or field direction. We demonstrate that there is no asymmetry in field dependences of the critical current and count rates in bend-free spirals, while in square spirals there exist an asymmetry in the field dependences of rates of dark and light counts. Invoking handedness (chirality) of the observed asymmetries and mapping the computed local absorption probability for photons and the local detection threshold, we identify areas in the bends where light counts at low photon energies originate from.

In the next section, we describe the manufacturing process of spiral structures and their characterization. We describe the experimental findings in a separate section, which is followed by the section with the theoretical model. Simulation results for the local absorption probability and discussion are presented in the last section.

## II. TECHNOLOGY AND EXPERIMENTAL DETAILS

Spiral structures were prepared from niobium nitride (NbN) films on sapphire substrates. We started by depositing a thin NbN film on an R-plane cut, one-side polished substrate via reactive magnetron sputtering of a pure Nb target in an argon and nitrogen atmosphere. Partial pressures of argon and nitrogen were $P_{Ar}= 1.9\times10^{-3}$ mbar and $P_{N2} = 4\times10^{-4}$ mbar, respectively. During deposition the substrate was placed without been thermally anchored on the surface of a holder which was placed on a heater plate. The plate was kept at a temperature of 850° C. The film thickness of $d$ = 4.8±0.2 nm was measured with a profilometer. A detailed description of the deposition process of NbN thin films can be found elsewhere [14]. We have chosen two designs of spirals for our experiment: normal spiral or Archimedean spiral (Fig. 1a) and the square spirals, sometimes also called Egyptian or Greek spiral (Fig. 1b). All spirals had one contact pad outside the spiral structure and one in its geometric center. The pad in the center was in the form of either a circle for the normal spiral or a square for square spiral with a diameter of 1.2 μm or sizes 1.3x1.3 μm$^2$, respectively. The geometric parameters of the spiral structures were measured with a scanning-electron microscope (SEM). The SEM images of spirals are shown in Fig. 1. All spirals, reported here, have a strip widths of $w$ = 110±5 nm and a strip spacing of 50 nm which both define a geometric filling factor of 70%. All bends in square spirals have nominally the same rounding radius $r$ = 71±5 nm at inner corners. The fabrication process of spiral structures includes three steps. To pattern our NbN films into spirals, we used electron-beam lithography over polymethyl methacrylate (PMMA) resist with a thickness of about 65 nm. The transfer of the image, created in the resist, was made by a subsequent milling with Ar ions at a pressure of $1.1\times10^{-4}$ mbar. We used an RF-plasma source from the firm Nordiko with a 100 mm diameter. The etching rate of NbN film ≈ 1.6 nm/min was achieved at an Ar flow of 4.8 sccm, 200 W applied RF power and 400 V ion-acceleration voltage.

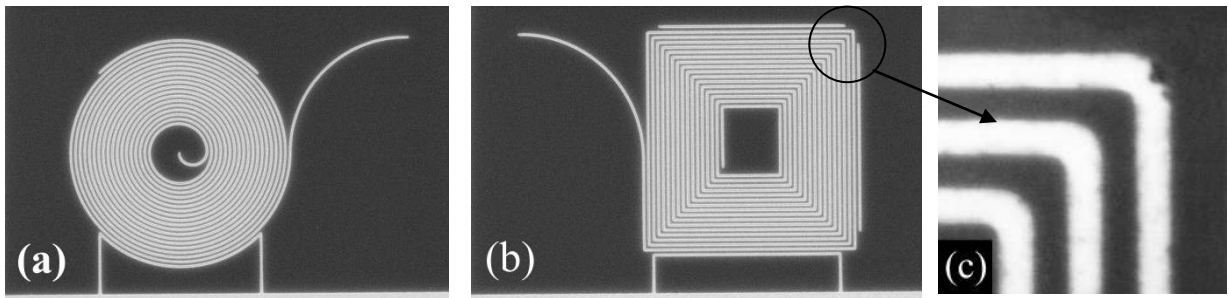

Fig. 1 SEM images of Archimedean (a) and square spiral structures (b). Dark color represents strips and surrounding fields from NbN film. The outer diameter of the Archimedean spiral is 7.3 μm. The size of the square spiral is 6.5 x 6.5 μm$^2$. (c) Corner rounding in bends of square spirals. Both the strip width and the separation are nominally 112 nm. Irregularities at the edge of the NbN field which surrounds the spiral structure do not affect the current ability of the spiral.

To lead bias current through, the spiral should be isolated from the top except for the central pad. Then a top electrode should be brought above the isolating layer. The schematics of the required contacting and isolating layers is shown in Fig. 2. At the second step, we made the isolating layer from aluminum nitride (AlN). A new PMMA layer was spun over the spiral structure and the ring with an outer diameter slightly larger than the outer diameter of the spiral and the inner diameter slightly smaller than the size of the central contact pad was opened. We further deposited 50 nm of AlN at room temperature by reactive magnetron sputtering of pure Al target in an argon/nitrogen atmosphere at partial pressure of argon and nitrogen $P_{Ar} = 3\times10^{-3}$ mbar and of $P_{N2} = 4.5\times10^{-3}$ mbar, respectively. After deposition, AlN from the central pad and surrounding of the spiral was removed in warm acetone via lift-off. The 50 nm layer of AlN reliably isolates the spiral structure from being short cut by the top electrode. The last step in fabrication of spiral specimens was processing of the top contact. To ensure a proper electrical contact to the spiral the top electrode must be at least two times thicker than the isolating layer. The top electrode was formed by e-beam lithography from a 100 nm thick Nb superconducting film which was deposited on top of the isolating layer by magnetron sputtering of pure Nb in an argon atmosphere at an argon pressure of $P_{Ar} = 5\times10^{-3}$ mbar. For e-beam lithography we used PMMA resist with a thickness of 120 nm.

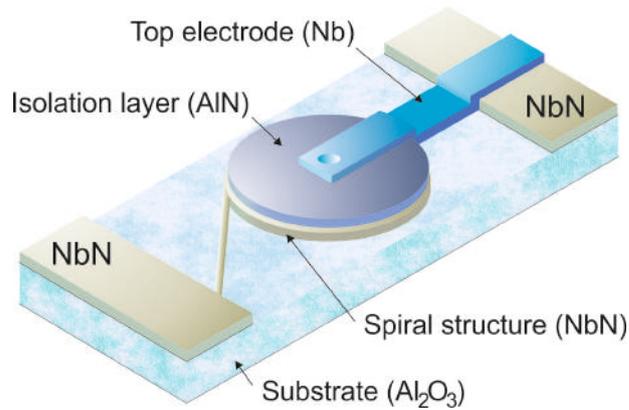

Fig. 2 (Color online) Schematics of the multilayer structure with the top electrode.

We measured the temperature dependence of the resistance of our spirals in the range from room temperature down to 4.2 K using standard four-probe technique. The critical temperature $T_C$ was defined as the lowest temperature at which a non-zero resistance could be measured. We found $T_C \approx 12$ K for all our samples with a variations from sample to sample of less than 0.3 K. Samples with a smaller strip width typically have a lower critical temperature [15, 16]. The current-voltage (CV) characteristics of the samples were measured in the current-bias mode at 4.2 K. The critical current $I_C$ of the spiral structures was associated with the well pronounced jump in the voltage from zero to a finite value corresponding to the normal state. The parameters of studied structures are listed in Table I.

Measurements in magnetic field were performed in the a home-made inset with a thermally isolated capsule for samples, where the temperature could be varied between 2 K and 15 K. The magnetic field up to 2 T was provided by a superconducting solenoid. Light from a monochromator was fed to the samples via a multimode optical fiber. We did not control light polarization which was slightly elliptical. Electrical readout was made via coaxial cable. In more details, the experimental setup was discussed elsewhere [7]. We checked that the dark count rate down to approximately $10^{-1}$ per second was current dependent. This eliminates electrical fluctuations as a source of dark counts. The critical current in magnetic field was measured in the voltage-bias mode via the long coaxial cable with an additional low pass filter. The critical current was defined as the maximum in the CV curves. At the critical current we typically found a rate of dark counts of $10^7$ to $10^8$ sec$^{-1}$ and a few microvolt voltage in excess to zero-resistance value.

TABLE I. Parameters of two typical spiral structures from NbN films: *RRR* – residual resistance ratio, i.e. the ratio of the resistance of the structures at room temperature to that at 20 K, $I_c(4.2 K)$ – critical current at 4.2 K.

| Type | *w* [nm] | *d* [nm] | *RRR* | $T_C$ [K] | $I_C$ (4.2 K) [µA] |
|---|---|---|---|---|---|
| Normal spiral | 104 | 4.8 | 0.98 | 11.7 | 35 |
| Square spiral | 112 | 4.8 | 0.97 | 11.9 | 36 |

### III. EXPERIMENTAL RESULTS

#### A. Critical current in magnetic field

We begin with the critical parameters of the superconducting state which provide scales for measured critical currents and applied fields. The depairing critical current in straight portions of the square spiral was computed in the framework of the standard Ginsburg-Landau (GL) approach with the Bardeen's temperature dependence and the correction for the extreme dirty limit as

$$I_C^{dep}(T) = \frac{4\sqrt{\pi}\left(e^{\gamma}\right)^2}{21\varsigma(3)\sqrt{3}} \frac{\beta_0^2 (k_B T_C)^{\frac{3}{2}}}{eR_S\sqrt{D\hbar}} w \left[1-\left(\frac{T}{T_C}\right)^2\right]^{3/2} K(T/T_C), \quad (1)$$

where $K(t) = 0.66 \times (3-t^5)^{0.5}$ is the analytical presentation of the correction [17], $R_S = 300$ Ω is the square resistance of our films at 20 K, $D = 5 \cdot 10^{-5}$ m² s⁻¹ is the typical diffusivity of normal electrons in our films [14, 18], $\beta_0 = 2.05$ is the ratio of the energy gap at zero temperature to $k_B T_C$ [19], $w$ is the width of strip in the spiral structure. For the square spiral with $w = 110$ nm and $T_C = 11.8$ K, we obtained a depairing critical current of 129 μA at $T = 4.2$ K. The second critical magnetic field $B_{C2} = 13.3$ T at 4.2 K was computed with the following expression

$$B_{C2}(T) = \frac{2\sqrt{2}\, k_B T_C}{\pi e D}\left[1-\frac{T}{T_C}\right]\left[1+\frac{T}{T_C}\right]^{\frac{1}{2}}. \quad (2)$$

It is expected that in a square spiral current crowding [2] at the inner corners of bends will reduce the measured critical current of the structure with respect to the critical current of the straight parts. External magnetic field induces screening current in bends. Depending on the field direction, the screening current may decrease or increase the local current density at the inner corner of a bend [11]. Fig. 3 shows a combination of field and current directions, which results in an increase of the local current density at inner corners. The same effect is achieved when the directions of both field and current are changed to opposite. We will call the direction combinations, which have such an effect on the current density, the state with left field-current symmetry. Two pictograms in the left box show two combinations with the left symmetry. The crosses or points in the circles denote two opposite directions of the magnetic field and the arrows – the directions of the bias current in the bend. The other two combinations will be named sates with right field-current symmetry. The corresponding two pictograms are shown in the right box. We will be using pictograms through the paper to relate experimental data on plots to specific combinations of field and current directions.

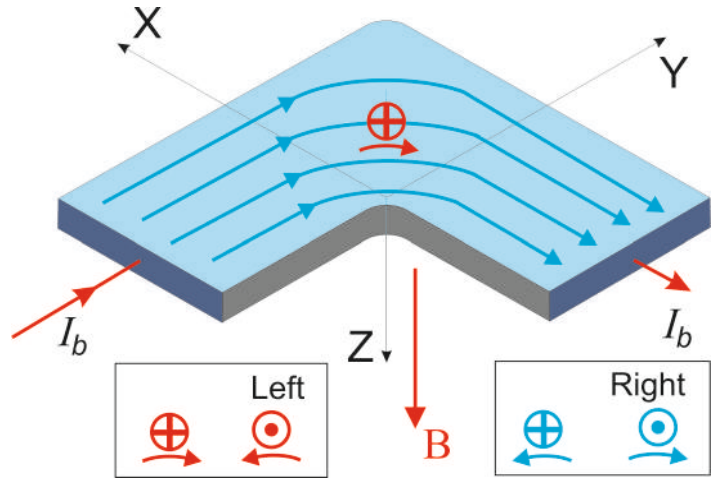

Fig 3. (Color online) Bend schematics and positive directions of external magnetic field (*B*) and bias current ($I_b$). These directions obey left field-current symmetry for which an increase in the magnetic field causes the increase in the superconducting current density at the inner corner of the bend. Pictograms in the left box denote two possible configurations which produce such effect. Pictograms in the right box denote configurations having right symmetry and, consequently, opposite effect on the current density at the inner corner. The pictogram in the bend corresponds to the positive directions of field and current shown in the figure. The inner corner has coordinates (0; 0; 0) in the system which is shown here and will be used through the paper.

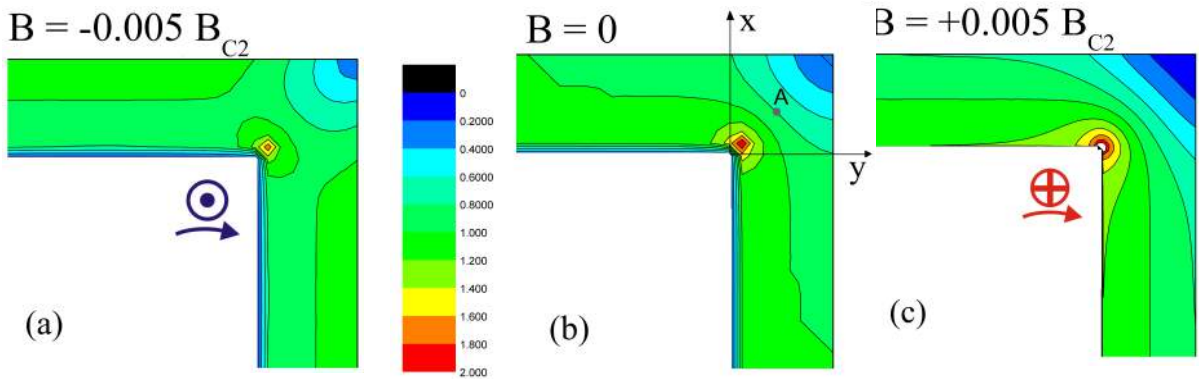

Fig. 4. (Color online) Density of superconducting current in the bend and adjacent straight parts without magnetic field (b) for the field $0.005B_{C2}$ with the positive (c) and negative (a) directions. The current density is normalized to its density far from the bend at $B = 0$. The current distribution was computed for the strip width of $20\xi$ where $\xi = 5$ nm is the coherence length for our NbN films [14, 18].

In order to visualize the expected effect of the magnetic field on the current density we computed in the framework of the GL formalism [20] the local density of superconducting current in a bend without magnetic field and a field of $B = +/- 0.005B_{C2} = 66$ mT. The results are shown in Fig. 4 as two-dimensional contour plots. The critical current is achieved when either the vortex barrier at the inner corner disappears [2, 4] or the local current density at the inner corner equals the depairing current density [21]. In both cases one expects the critical current to decrease with increasing field for the left field-current symmetry and to increase for the right field-current symmetry.

The critical current was measured for both, field and current, directions as a function of the magnetic field. The results are shown in Fig. 5 for the square spiral (a) and for the normal spiral (b). The square spiral demonstrates dependences expected for a single bend. For combinations with the right symmetry, the critical current increases with the field, reaches a maximum at $B_{max} = 44$ mT and further decreases. For combinations with the left symmetry, the critical current linearly decreases with magnetic field. This effect was already reported for separate bends [12, 13]. Simultaneous change of current and field directions mirrors the $I_C(B)$ curves with respect to $B = 0$ line. An archimedean spiral does not have any asymmetry of the critical current in magnetic field. Assuming that all bends in the square spirals are identical, we apply the analysis of Ref. 13 to find via linear extrapolation of the field dependence for the right symmetry the critical current in the straight parts $I_{C0} = 42$ µA and the reduction factor $R = I_C/I_{C0} = 0.86$ due to current crowding. The critical current in straight parts of the spiral is less than the computed depairing current. The difference is within the range found for nanowires with similar stoichiometry [14]. The self-field that is produced by the critical current in the middle part of our spiral structures is less than 0.1 mT and is from two to three times larger than the local earth magnetic field and almost two orders of magnitude less than typical $B_{max}$ values.

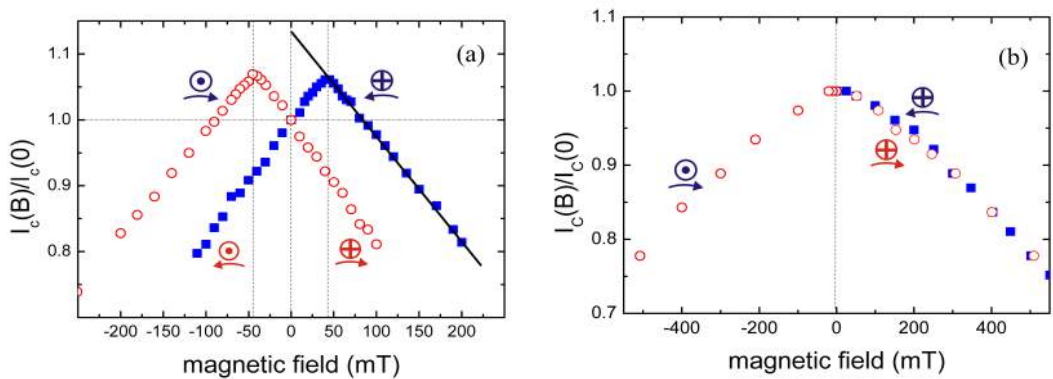

Fig. 5. (Color online) (a) Critical current of the square spiral in magnetic field for positive (open symbols) and negative (closed symbols) directions. Pictograms depict combination of the field and current directions for each section of the plot. Solid straight line extrapolates linear decrease of the critical current in the right symmetry to the zero field. Vertical dashed lines show zero field and positions of the maxima on the field axis. (b) Critical current of the Archimedean spiral for different current directions. The same convention is used to mark symmetries and current directions.

## B. Dark counts

The rate of dark counts in square spiral is not symmetric with respect to the direction of either magnetic field or current. The minimum in the magnetic field dependence of the dark-count rate (DCR) appears for the same right field-current symmetry as the maximum in the field dependence of the critical current. This is illustrated in Fig. 6 where DCR is plotted as function of the magnetic field for two opposite directions of the bias current. Like the critical current, DCR is invariant for changing simultaneously the directions of both field and current. Noticeably, the minimum in DCR occurs at a field of approximately 25 mT which is smaller than the field corresponding to the maximum in the critical current. Increasing the bias current does not affect the position of the minimum in DCR but makes it more pronounced and sharp (Fig. 6 (b)). In Archimedean spirals DCR was found symmetric with respect to field and current directions for any fields and currents.

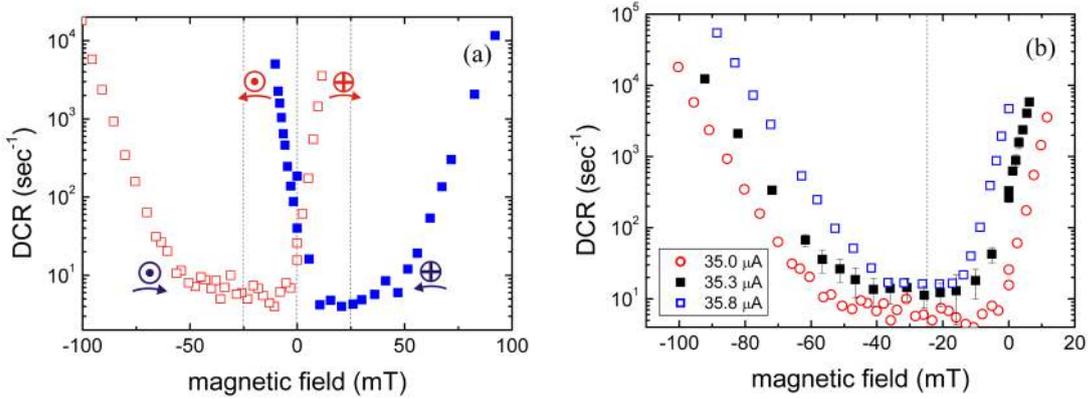

Fig. 6. (Color online) (a) Rate of dark counts in magnetic field for two directions of the bias current with the magnitude 35 μA. DCR for positive current direction is shown with open symbols and for negative direction with closed symbols. Pictograms depict combination of the field and current directions for each section of the plot. Dashed vertical lines are to guide the eyes, they show field positions of the minima in DCR and zero field. (b) Magnetic field dependencies of DCR for different positive bias currents. Values of the bias current are specified in the legend. Vertical dashed line shows the location of the minimum on the field axis.

## C. Photon counts

Although, similar to dark count rate, the rate of light counts exhibits an asymmetry separately with respect to field and current directions, the effect of field and current appears more complicated. First, the strength of asymmetry in the photon count rate (PCR) depends on the photon energy. Fig. 7 shows the rate of photon counts in magnetic field for three wavelengths 1400 (a), 800 (b) and 500 nm (c) and different bias currents. As it has been reported earlier [7], the change in the PCR produced by the same magnetic field decreases with the decrease in the wavelength and varies from two orders of magnitude for the wavelength $\lambda = 1400$ nm to tens of percent for $\lambda = 500$ nm. An increase in the bias current reduces the range of changes for each wavelength. Although, like DCR, PCR is also invariant for changing simultaneously the directions of both field and current, the asymmetry in PCR qualitatively differs from the asymmetry in DCR. Remarkable, that the minimum in PCR appears for the left field-current symmetry (Fig. 7 (a)) and not for the right symmetry as it is for DCR. In other words, for the same current direction the minimum in the PCR is shifted in the opposite direction on the field axis as compared to DCR. We will discuss this counterintuitive behavior at the end of this section. The absolute value of the field at the minima in PCR for $\lambda = 1400$ nm is approximately 17 mT which is less than the field value at the DCR minima. The asymmetry in PCR is more pronounced for large wavelengths and small currents and disappears completely for wavelengths smaller than approximately 600 nm. For the wavelength 800 nm and the bias current 27 µA the asymmetry is still distinguishable (Fig. 7 (b)) while it is already hard to see at a bias current of 29 µA. Within our experimental accuracy we did not find any asymmetry for the wavelength of 500 nm (Fig. 7 (c)). The upturn in the plots for the bias current of 32.4 µA occurs when the critical current in the field decreases down to the bias current. We did not find any asymmetry in the PCR dependences on the magnetic field for the Archimedean spirals.

The effect of an external magnetic field on the critical current and rates of dark and light counts, which we described above, makes it possible to come to certain conclusions without invoking the qualitative microscopic analysis. Excluding large single defect somewhere at the strip edge in the square spiral structure, we have to accept that any dark or light count events whose rate is asymmetric with respect to the direction of either field or current alone come from the bends in the structure. For the critical current this was already justified in experiments with separate bends [12, 13]. The maximum of the experimental critical current in magnetic field is achieved when increasing critical current in bends equals decreasing critical current in straight strips. For count events in a straight strip, any microscopic model would predict symmetric field or current dependencies of corresponding rates because the strip itself and the absorption probability for photons are both symmetric over the midline of the stripe and the distributions of the current density and magnetic field in the strip have even and odd symmetry, respectively, with respect to the midline. Hence, when the field or current directions change this will not affect the critical current and count rates which should remain unchanged. The square spiral consists from straight strips and bends. Therefore, any asymmetry may come from bends only. Indeed, in Archimedean spirals where no sharp corners are present and the rounding radius of the spiral is much larger than the strip width we did not observe any asymmetry. Furthermore, a weak asymmetry in the dark count rate with respect to the field direction has been recently observed in meanders [5] which contain turns with different symmetry in small but non-equal numbers. Here, the net asymmetry may arise from a slight difference between geometrical shapes of individual turns.

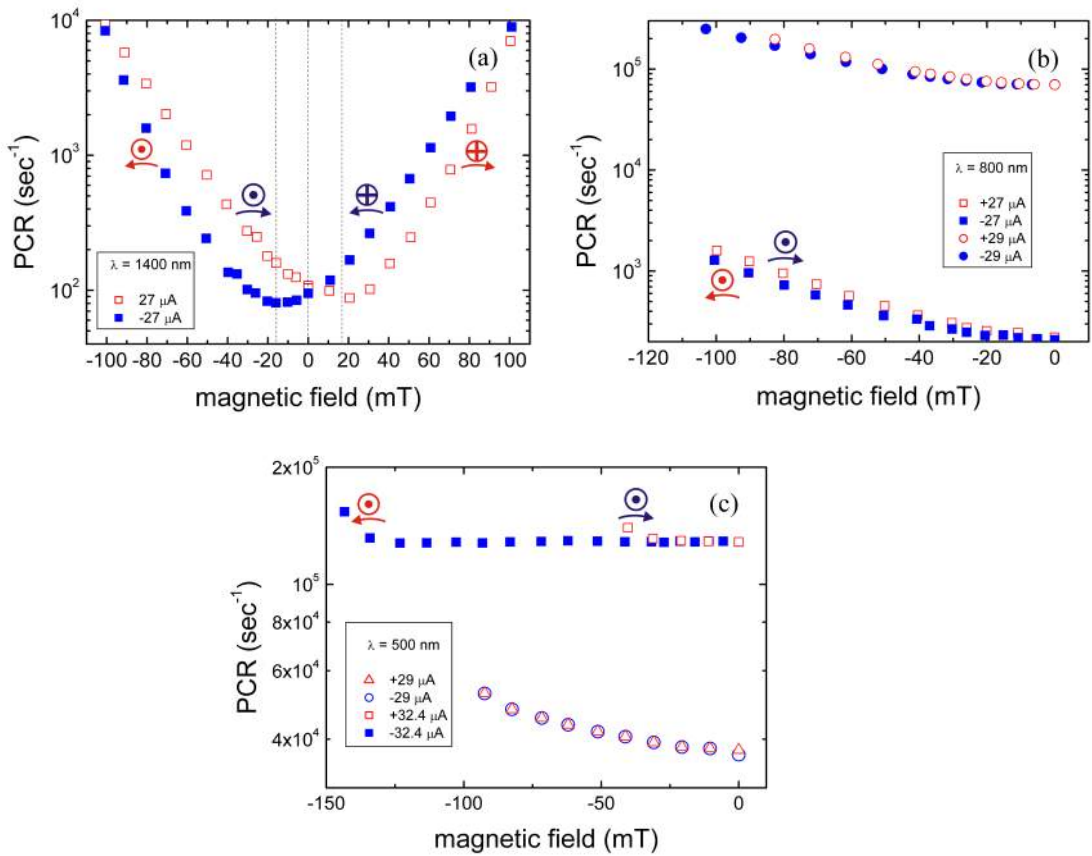

Fig. 7. (Color online) (a) Rate of light counts in magnetic field for different wavelengths: 1400 nm (a), 800 nm (b) and 500 nm (c). Magnitudes and directions of the bias current are specified in the legends. Conventionally, PCR for positive current direction are shown with red symbols, blue symbols correspond to the negative direction. Vertical dashed lines in the panel (a) guide the eyes to the locations of the minima in PCR on the field axis. Pictograms depict combinations of the field and current directions for each section of plots.

The phenomenological explanation of the asymmetry in DCR is straightforward. For the left symmetry of the field-current directions the field increases the current density at the inner corner of the bend. This reduces the potential barrier for vortices entering the bend from the inner corner and, correspondingly, increases the rate of dark counts. The field applied in the right symmetry decreases the current density at the inner corner and decreases the count rate. When the field in the right symmetry further grows, the current density at the outer edges of straight strips increases and reduces the barrier for anti-vortices. At some field they begin to dominate the net count rate and DCR starts to increase. Somewhere at an intermediate magnetic field, the count rate drops to minimum. Since the rate of events from straight segments is symmetric with respect to the field direction and have a different field dependence as compared to the rates from bends, the net rate may have a minimum at a field smaller than the field that maximizes the critical current.

Intuitively one would expect the same kind of non-symmetry for the rate of light counts. However, this expectation silently postulates that light and dark counts undergo the same microscopic scenario. This is not necessarily the case. Recently, it has been found that the microscopic scenario of photon detection as well as the detection efficiency may differ locally [8, 10, 22]. Let us consider the point on the common bisector of both corners in the bend close to its midline, e.g. point A in Fig. 4b. In the absence of an external field, the current density at the selected point is less than at the inner corner. The photon which is absorbed at this point creates a hot-spot. The hollow in the order parameter forces the supercurrent to flow around and increases velocity of the condensate at the edges of the hot-spot [8]. An external field in the left symmetry will decrease the current density locally around the hot-spot. The photon is counted as light event if either the velocity locally reaches the critical value and an VAP appears or a vortex enters the hot-spot from any side and then moves to the opposite one. An increasing field either disables VAP appearance or increases the barrier for vortex around the hot-spot. The local photon count rate decreases either way. Obviously, the field applied in the opposite direction causes an increase in the local PCR. The net effect crucially depends on the distribution of the absorption probability in the bend and the bias current. In the next section we show that two-dimensional GL model qualitatively explains different asymmetries in dark and light count rates.

## IV. THEORETICAL MODEL

We first discuss critical currents in square spiral structures. We found the critical current, i.e. the current at which the superconducting state becomes non-stable, from the numerical solution of the GL equations [20] in the geometry shown in the inset in Fig. 8. We considered separately bends (B) where we neglect rounding and straight segments (A) of the spiral with edge defects. The results are presented in Fig. 8 separately for bends and for straight segments. We found that the maximum of the critical current in the bend (closed symbols in Fig. 8) should occur for the right symmetry at $B \approx 0.02\,B_{C2} \approx 260$ mT. This is almost twice as large as the value obtained with the London model (Eq. 17 in Ref. 11) for a sharp 90° bend. Taking into account nominal rounding of inner corners in bends of our structures $r/w = 0.65$ and assuming that all bends are identical, we expect in the framework of the London model a reduction factor of $R = 0.75$ (Fig. 14 in Ref. 2) for the critical current in bends and the maximum in the experimental critical current at $B_{max} = 65$ mT (Eq. 17 in Ref. 11). Our experimental values $R = 0.86$ and $B_{max} = 44$ mT (Fig. 5) are reasonably close to predictions of the London model. Moreover, for our experimental reduction factor $R = 0.86$ the London model predicts $B_{max} = 38$ mT which even better corresponds to our experimental value $B_{max} = 44$ mT. We attribute the remaining difference between the experimental reduction factor and the factor, which is expected for a nominal rounding in our structures, to geometrical non-uniformities of the strip edges. Such non-uniformities typically appear as a result of ion etching [23]. They slightly decrease the effective width of strips in straight segments and increase the effective rounding radius of inner corners in bends.

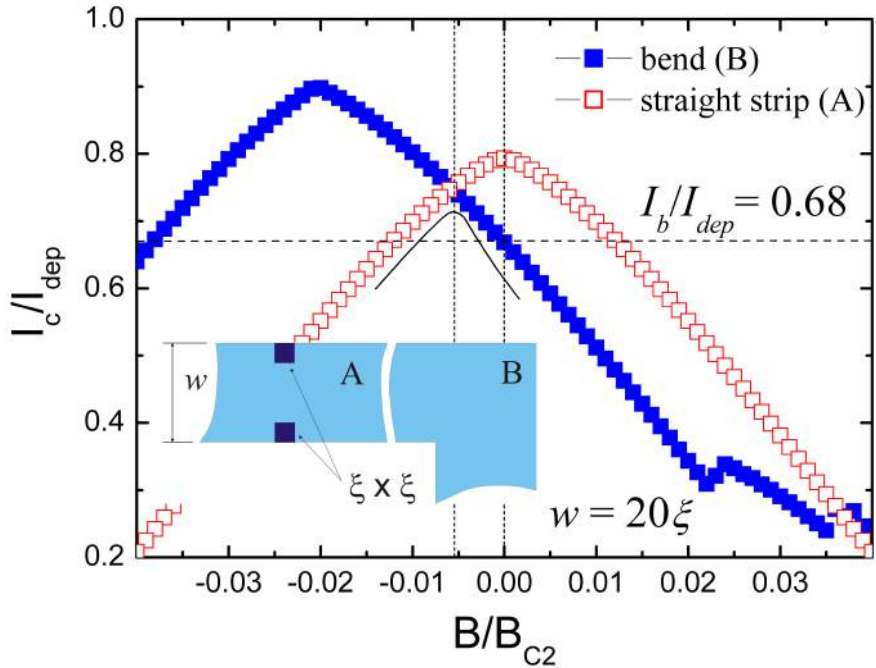

Fig. 8. (Color online) Relative critical current at different magnetic fields for the sharp bend in the strip with the width $w = 20\xi$ (closed symbols) and for the straight part of such strip with defects (open symbols). The inset

shows the geometry used for modeling: A - straight strip with defects (not in scale); B – sharp bend. Solid line shows currents which appear as critical currents of a structure consisting from bends and straight strips. Vertical lines guide eyes to zero field and to the expected maximum in the critical current of the structure. Horizontal line shows the bias current which was applied to measure DCR (Fig 6(a)).

To model within the GL approach the reduction of the critical current in straight segments of the spiral we introduced two identical defects at the opposite edges of the straight strip (Part A of the inset in Fig. 8). Both defects represent a local suppression of the order parameter in an area $\xi \times \xi$ adjacent to the strip edge. They reduce the critical current in the straight strip to 80% of the depairing critical current in the strip of the same width without defects. The dependence of the critical current in the strip with defects on the magnetic field is shown in Fig. 8 with open symbols. The critical current of the whole structure, which is comprised from bends and straight strips, will be limited to the smallest value out of critical currents of these two components. The solid line in Fig 8 shows the path which the critical current of the complete structure follows with varying magnetic field. In accordance to our experimental data, the maximum in the critical current occurs at $\approx$ 50 mT. It corresponds to the intersection of curves for the bend and for the straight strip. We understand that our experimental dependence of the critical current in the spiral on magnetic field can be modeled by a different set of rounding radius in the bend and the size of defects in straight parts, e.g. smaller defects and a larger rounding radius. Since we cannot visualize defects, the set which we used to obtain our model currents is rather arbitrary. However, we do not anticipate any effects of this choice on the asymmetry in light and dark count rates in magnetic fields.

Although dark counts are generated in all parts of the spiral structure, the rates per unit length (local DCR) depend crucially on the ratio between the local critical current and the bias current. The local DCR is proportional to $\exp(-\delta F/(k_B T))$ where $\delta F$ is the local height of the barrier for vortex entry. In the framework of the London model for large bias currents $I_b \leq I_C$ the barrier scales with the difference between the local critical current and the bias current $\delta F \propto \delta I = I_C(B) - I_b$ [9]. Therefore for any bias current the local DCR in bends will be much higher than the local DCR in straight strips. The total DCR in the spiral will depend on the relative weight of bends and straight strips. However, since the local DCR in strips is symmetric with respect to the direction of the magnetic field, the presence of any asymmetry in the magnetic field dependence of the total DCR ensures the non-vanishing contribution of bends to generation of dark counts. To compare our experimental results with the model calculations quantitatively, we identify the depairing current in the GL model with the critical current in the straight strips in zero magnetic field. Taking into account 20% reduction of the model critical current in straight strips due to defects, we arrive at $I_b/I_{dep} = 0.68$ for the bias current which we used for DCR measurements. This relative bias current is marked with the straight dashed line in Fig. 8. If bends noticeably contribute to the total DCR one would expect a minimum in total DCR at a field close to our experimental value of $B_{max}$. Plots in Fig. 6 (b) confirm this expectation. There is a minimum in DCR around -30 mT. Slopes of the dark count rate versus magnetic field are different for fields with right symmetry at $B < -50$ mT (less dark counts from bends) and for fields with left symmetry at $B > -10$ mT (more dark counts from bends). Different slopes correspond to different weights of bends and straight segments in the total DCR. Since the bends do not dominate in the total DCR at all magnetic fields the minimum in the DCR($B$) dependence does not coincide with the maximum in the $I_C(B)$ dependence (compare Fig. 5(a) and Fig. 6(a)).

Accepting the thermally activated vortex crossing as the dominant photon-detection mechanism one would expect for the photon count rate the same type of asymmetry as for the dark count rate. Indeed, the vortex should enter the superconductor and hot spot via the weakest place, i.e. the place where the current density/supervelocity is maximal. This can be the inner corner of the bend, especially when the hot spot is located close to it and the field of the left symmetry favours the entrance of a vortex with the same polarity as in the case of dark counts. However, contrary to dark count events, in the light-count scenario the vortex should also exit the hot spot. The local value of the order parameter inside the hot spot $\Delta$ is less than the equilibrium value $\Delta_{eq}$ outside of the hot spot. If the relative local decrease of the order parameter is small $\delta = (\Delta_{eq}-\Delta)/\Delta_{eq} \ll 1$ the hot spot cannot pin the vortex and vortex passes freely across the strip. When the relative decrease is large the hot spot pins the vortex and prevents the light count. Whichever of these two occurrences holds for particular order parameter in the hot spot, depends on the bias current. In Fig. 9 we plot the current at which the vortex leaves the hot spot as a function of the location of the hot spot in the bend. We call this current the detection current $I_{det}$ since it ensures the light count. Calculations are made in the framework of the modified hot-spot model [24] with the radius of the hot spot $R = 5\xi$ and the relative reduction of the order parameter $\delta = 0.4$.

One can see that close to the inner corner there is an area in the bend where the small field of the left symmetry increases $I_{det}$ while the field of the right symmetry decreases $I_{det}$. This area is marked schematically with grey colour in the inset in Fig. 9(a). Positions on the cut through this area at $y = w/4$ are encircled in Fig. 9(a). Here, they span over the interval of bias currents $0.39 I_{dep} < I_b < 0.44 I_{dep}$. For any relative bias current within this interval, only the part of the encircled area where $I_{det} < I_b$ provides light counts. This active part decreases if a small positive (left symmetry) magnetic field is applied and increases if magnetic field is negative (right symmetry). Because of the uniform and constant photon flux the light count rate is proportional to the area which is collecting photons, the active part will deliver PCR with the asymmetry which we observed experimentally. This asymmetry is inverted with respect to the "normal" asymmetry of dark count rate.

Note that in the active part, negative magnetic field favours *exit* from the hot spot of those vortices, which have entered the hot spot from the side of the inner corner. At the outer edge of the hot spot, far from the inner corner the negative magnetic field increases locally the current density (and supervelocity) and decreases the energy barrier for vortex exit. The inverted asymmetry exists only in small magnetic fields. This corresponds to our experimental observation. PCR at $\lambda = 1400$ nm becomes symmetric for fields larger than 100 mT (Fig. 7(a)). The inverted asymmetry disappears for $\delta = (\Delta_{eq}-\Delta)/\Delta_{eq} > 0.5$ which corresponds photons with higher energy. In this case, $I_{det}$ in the bend and in the straight strip are practically equal. The effect is also absent when the hot spot has no vortex pinning ability, e.g. when $\delta < 0.3$. At bias currents larger than $I_{det}$ in straight strips, light counts come mostly from straight strips and any asymmetry in PCR disappears.

Under the same conventions as in the case of dark counts, we find that the current 27 μA, which we used to measured PCR at $\lambda = 1400$ nm, corresponds to the model-relevant relative bias current $I_b = 0.5\ I_{dep}$. This current is at the boundary of the current interval where the inverted effect exists. However, since our choice of the rounding radius and the size of defects for the model dependence of the critical current on the field is not unique (see discussion above), the relative bias current may have different value. In other words, we are not able within the present model to estimate numerically the relative weight of bends in the total rate of light counts.

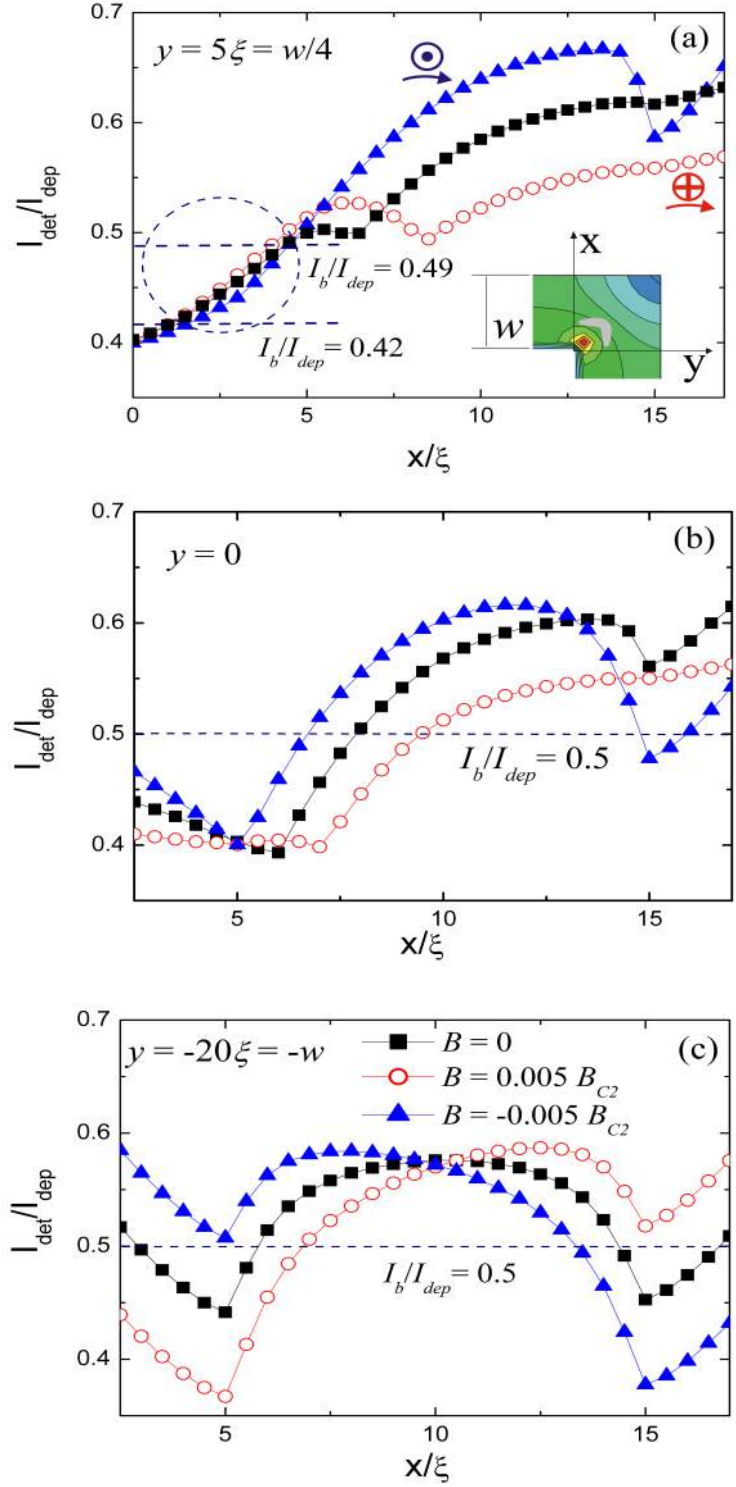

Fig. 9. (Color online) Detection current as function of the positions ($x$) of the hot spot at three different distances ($y = w/4$ (a); $y = 0$ (b); $y = -w$ (c)) from the inner corner of the bend without magnetic field and for the magnetic field $B = 0.005 B_{C2}$ with different symmetries. Positive magnetic field corresponds to the left symmetry. Coordinate system is shown in the inset in the panel (a). The inset in panel (a) sketches the bend and the area where $I_{det}$ is increased/decreased by small magnetic field of the left/right symmetry. The cut through this area at $y = w/4$ is marked with red dashed circle in the panel (a). Horizontal straight lines in panel (a) show boundaries for bias currents within which the effect exists. Horizontal line $I_b = 0.5 I_{dep}$ in panels (a) and (b) shows the nominal value of the bias current which was used to measure PCR for the wavelength 1400 nm (Fig. 7(a)).

We are aware that the analysis [25] based on the solution of the time-dependent GL equations has shown that light counts generated by bends differ from those originating from straight strips. More specific, the overall duration of a PCR voltage pulse is smaller when the count originates from the bend and, at small bias currents, the amplitude of a PCR pulse from the bend is smaller than the amplitude of the pulse from the straight strip. Recent work which invokes the same theoretical approach [26] predicts a similar difference between the amplitudes of PCR voltage pulses originating from bends and strait sstrips with constrictions. With our spirals we observe PCR and DCR pulses with equal mean amplitudes and an amplitude spread, which is much narrower than the both models predict. The time resolution of the present experiment (approximately 100 ps) does no allow us to resolve the passage of a kinematic vortices. Early experiments on meander structures which include $180^o$ turnarounds had demonstrated a difference between mean amplitudes of PCR and DCR pulses and also a decrease in the mean amplitude of PCR pulses with an increase of the photon energy [27]. These early observations contradict to the results of both models. The reason of this discrepancy is not clear at the time. Anyway, like in the case of dark counts, the asymmetry in PCR itself ensures a noticeable contribution of the bends to the total rate of light counts.
We believe that the time-dependent GL equation alone cannot provide correct (quantitative) description of this problem. Without solving coupled GL and kinetic equations it is not possible to state unambiguously whether a passage of a single Abrikosov vortex or series of kinematic vortices leaves enough heat to create a normal resistive domain. Instead of the kinetic equation authors of the both models solved heat conductance equation. This approach is only qualitatively valid, because the time for vortex nucleation is smaller than the electron-electron inelastic relaxation time and the usage of the effective temperature is not validated. Furthermore, in Ref. 25 the coefficient which describes heat transfer from electrons to phonons was larger than its typical value in NbN films that resulted in too small local heating. With more realistic value of this coefficient [26] one finds that the photon absorbed near the bend creates normal domain at a smaller current than the current required to generate light count in the straight segment.

## V. DISCUSSION AND CONCLUSION

Theoretical considerations presented in the previous section are based on a relatively simple microscopic model of the hot spot [24]. The actual profile of the order parameter in the hot spot may differ from the assumption of this model which will quantitatively influence the pinning ability of a hot spot and the detection current. Furthermore, we cannot relate precisely the bias current in the experiment to the particular relative bias current in the GL model. Therefore the contribution to the net PCR from different parts of the bend remains largely undefined. For the bias current $I_b = 0.5 I_{dep}$ the hot-spot positions around the geometric border between the bend and the straight strip ($y = 0$, Fig. 9 (b)) contribute to the net PCR with even larger area than the central part of the bend. The positions at $0 < x/\xi < 5$ contribute with inverted asymmetry while positions at $5 < x/\xi < 10$ – with normal asymmetry. The net DCR from these areas may be well symmetric. The straight strips contribute symmetrically to PCR at any bias current. Therefore, they smear out the shift of the minimum in PCR to either side. This interplay of light counts from different parts of the spiral structure is further modified by the probability of photon absorption.

We computed this probability for plane waves with three different polarizations. We identify the probability of the photon absorption at a particular location with the density of the high-frequency current which is induced in the structure by the plane wave at normal incidence. Simulations are carried out with the software COMSOL [28] which implements the Finite-Element method. To verify that the simulation results are not affected by numerical instabilities or similar problems, we compare the results obtained with COMSOL to similar simulations done with the software Lumerical [29] which is based on the Finite-Difference Time-Domain method. The results provided by these two techniques almost coincide. The COMSOL software solves numerically Maxwell's equations in the frequency domain. The sample is modelled by its specific geometry and specified by its frequency dependent dielectric function. The calculations lead to an accurate theoretical treatment of the problem and the results automatically include surface plasmons if they are excited. Therefore no further separate analysis of surface plasmons is necessary. In all simulations, a maximum mesh size of 7 nm and a complex dielectric function for our NbN films in the normal state [30] are used. By comparing simulation results for a separate strip with a bend with the results for the whole spiral structure, we checked that there is no cross-talk between adjacent stripes via evanescent fields. The results are shown in Fig. 10 as gray-scale plots. They present current density in the equatorial surface of the structure at the frequency of the incident wave. For polarizations along *x* or *y* axis (Fig. 10(c)) approximately half of the bend is active in absorbing photons. The absorption probability is evenly distributed between hot-spot positions providing count rates with different asymmetry. The polarization at 45 degrees will be seen differently by adjacent bends. The two possibilities are shown in panels (a) and (b) in Fig. 10. The polarization along the bisector (Fig. 10 (a)) delivers more photons to positions providing normal asymmetry while photons with perpendicular polarization (Fig. 10 (b)) will be stronger absorbed at positions providing inverted asymmetry. The net effect of the absorption probability on the asymmetry in PCR seems to be very weak. Therefore, we did not attempt to convolve the map of the absorption probability with the map of detection currents.

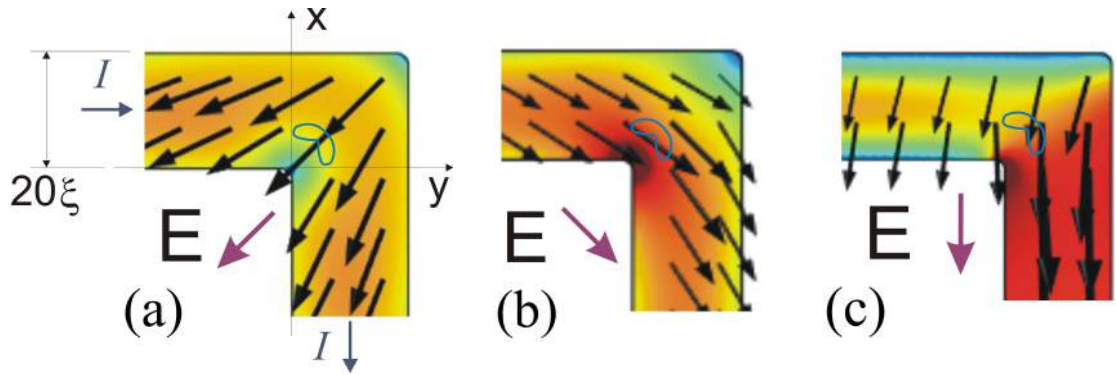

Fig. 10. (Color online) Computed probability of photon absorption in the bend and adjacent portions of the straight strips for different polarizations of incident photons with the wavelength 1400 nm. Polarization is shown with the arrow outside the bend. Black arrows show directions of currents inside the bend. Red (dark) color corresponds to the largest local probability. Blue curved lines encircle the area which deliver PCR with inverted asymmetry.

In this paper, we have demonstrated that in structures consisting from straight strips and bends with only one type of symmetry with respect to the direction of fields and currents exists an asymmetry in count rates for light and dark count events with respect to the direction of an external magnetic field and, separately, to the direction of the current. We have associated this asymmetry with the asymmetry in current crowding in the bends. Applying a simplified microscopic GL model we have shown that count events which provide asymmetry originate from bends while the rate of events coming from straight strips remains symmetric with respect to field and current. The microscopic scenario of the light count event with intermediate pinning of the magnetic vortex in the hot-spot explains the faint effect of the inverted asymmetry for low-energy photons at small fields and currents. We have shown that, in accordance with the predictions of the theoretical model, any asymmetry in the rate of light counts disappears at large magnetic fields and currents.

**AKCNOWLEDGEMENTS**

D.Yu.V. acknowledges support from the Ministry of Education and Science of the Russian Federation (State contract No 02.B.49.21.0003) and from the Russian Foundation for Basic Research (project 15-42-02365/15); I.C., K.D. and N.B. acknowledge support from Karlsruhe School of Optics & Photonics of Karlsruhe Institute of Technology.